\documentstyle[preprint,aps]{revtex}

\input epsf

\begin{document}
\draft
\preprint{}

\def\r{{\bf r}}   \def\q{{\bf q}}   \def\df{\delta\phi}
\def\udp{u^{\prime\prime}}
\def\ufp{u^{\prime\prime\prime\prime}}    \def\bu{\bar u}
\def\du{\delta u}
\def\XL{{x\over\sqrt{2}\lambda}}

\title{Counter-ion release and electrostatic adsorption}

\author{P. Sens \& J.-F. Joanny}
\address{Institut Charles Sadron, 6 rue Boussingault, 
67083 Strasbourg Cedex, France\\Email:sens@ics.u-strasbg.fr}
\date{\today}
\maketitle
\begin{abstract}

The effective charge of a rigid polyelectrolyte (PE) approaching an
oppositely charged surface is studied. The cases of a weak (annealed)
and strongly charged PE with condensed counterions (such as DNA) are
discussed. In the most interesting case of the
adsorption onto a substrate of low dielectric constant (such as a
lipid membrane or a mica sheet) the condensed counterions {\it are
not always} released as the PE approaches the substrate, because of the
major importance of the image charge effect. For the adsorption onto
a surface with freely moving charges, the image charge effect becomes
less important and full release is often expected.

\end{abstract}

%
%  Include PACS numbers and Journal to which paper will be submitted.
%
\pacs{PACS numbers: 61.25.Hq, 68.10.Cr, 87.14.Gg \\
submitted to {\em Phys. Rev. Lett.} on 01/19/00}
%
%  Text of the paper
%

\narrowtext

A deep understanding of the adsorption of DNA or other charged 
biomolecules onto oppositely charged membranes is of fundamental 
importance to understand many key physiological processes, and many 
experimental studies have approached this problem from very different 
viewpoints\cite{ref1}, strongly motivated by applications to gene 
therapy\cite{gt}. On the theoretical side, the adsorption and the 
interaction of charged rods on a charged surface have been studied within 
the Debye-Huckel approximation in different situations\cite{ref2}.
In this work, we want to focus on the effective charge of the adsorbed 
macromolecule. We calculate the  attraction energy between an
infinitely long, charged cylinder
(rigid polyelectrolyte or PE) parallel to an oppositely charged plane, as a
function of their distance
$h$. The energy variation leads to the determination of the equilibrium 
charge density of the rod as a function of $h$. This (effective) charge density can be interpreted 
in terms of the release of condensed
counterions for highly charged PE such as DNA, which
are beyond the Manning condensation threshold, or in terms of the
recombination of ionized charges on the rod for weak (annealed) PE.
We show that in contrast to what would be naively expected, the full 
release of the counterions condensed onto an highly charged rod is 
not always observed in the vicinity of an oppositely charged surface.

The adsorption energy is derived by perturbating the Gouy-Chapmann
solution  of the Poisson-Boltzmann equation in a planar
geometry. The perturbative treatment is strictly speaking valid only for
low linear charge
densities $\tau$ of the rod $\l_B\tau\ll1$, where $l_B$ is the Bjerrum length
$l_B=e^2/(4\pi \epsilon)$
(all energies are in $k_BT$ units). However, the physical picture
which emerges from
this calculation leads to qualitative statements concerning
highly charged PE as well.
We discuss the adsorption free energy on a substrate of low dielectric
constant
 with respect to water ($\epsilon_w=80$), which is of most practical
importance
 in biology related problem (adsorption of DNA on a lipid membrane
for which $\epsilon_{lp}\simeq2$\cite{lipid}) and other situations
(adsorption onto the mica
surfaces of an SFA: $\epsilon_{mc}\simeq 6$\cite{mica}). We have checked that 
the case of a membrane of thickness $l=50A$ and dielectric constant 
$\epsilon_{lp}=2$ does not show quantitative differences with the 
present situation ($l=\infty$ and $\epsilon_{lp}=0$). Finally, we also
 study the case where the charges on the plane are free to adjust 
 to the field created by the rod, a situation of great
interest for fluid interfaces such as biological lipid membranes.

The electrostatic potential $\phi^{(0)}$ near a charged wall of density 
$\sigma>0$ (or equivalently with a Gouy-Chapmann length $\lambda\equiv 1/(2\pi l_B
\sigma)$), in a salt solution of average concentration $n_0$ (or
Debye length $\kappa^{-1}$ with $\kappa^2\equiv8\pi l_B n_0$)
satisfies the Poisson-Boltzmann (PB) equation\cite{andelman}. In the low salt limit
$\kappa\lambda\ll1$, which is the case discussed in this paper, the
potential near the wall ($\kappa z\ll1$) follows the Gouy-Chapmann (G-C) 
law\cite{safran}:
\begin{equation}
\phi^{(0)}=2\log{\kappa(\lambda+z)\over 2}\quad n^{(0)}(z)={1\over
2\pi l_B(\lambda+z)^2}
\label{GC}
\end{equation}
where $z$ is the coordinate normal to the wall ($z=0$ at the
interface), $n^{(0)}$ is the density of (negative) counterions (or c-i)
near the wall. The Gouy-Chapmann solution predicts a dense counterion
layer (the G-C layer) of thickness $\lambda$ containing
a finite fraction of the c-i, followed by a diffuse c-i region.
The local screening
length in the G-C layer is small:
$L_\kappa\simeq\lambda/\sqrt{2}$,  while the screening length
of the diffuse region is self-similar: $L_\kappa\simeq
z/\sqrt{2}$. For distances larger than the Debye length:
$z>\kappa^{-1}$, the electrostatic potential decreases exponentially.

The potential variation $\delta\phi$ due to a negatively charge rod (of linear
charge density $-\tau$) located at an altitude $z=h$ above the plane
is calculated by a linear expansion of the G-C theory,
provided that the perturbation is small. A small perturbation of Eq.(\ref{GC}): $\delta
n\ll n_{}^{(0)}$ (note that $\delta n$ is an algebraic average taking
into account the sign of the free charges) supposes that the
perturbation potential $\delta\phi$ is smaller than unity. The
calculation of the potential is carried out in Fourier space for
the coordinate $x$ parallel to the wall and perpendicular to the rod:
$\tilde f_q=\int dx e^{i q
x}f(x)$. The linearized PB equation and the boundary conditions are:
\begin{eqnarray}
\partial^2_z\delta\tilde\phi=\left(q^2+
{2\over(z+\lambda)^2}\right)\delta\tilde\phi\qquad
\partial_z\delta\tilde\phi|_{z=0}=0\cr
 {\rm and}\quad
\partial_z\delta\tilde\phi|_{z=h_+}-\partial_z\delta\tilde\phi
|_{z=h_-}=4\pi l_B\tau
\label{PPB}
\end{eqnarray}with natural boundary conditions for
$z\rightarrow\infty$

From the perturbed potential $\delta\phi(x,z)$, the free
energy due to the presence of the rod can be calculated by a charging process of the rod: $\delta{\cal
F}=\int_0^\tau \left(\phi^{(0)}_{z=h}+\delta\phi_{z=h}\right)d\tau$.
The first part of the integral gives the interaction energy between
the rod and the charged plane and the unperturbed G-C counterion layer,
while the second part represents the self-energy of the rod, and its
interaction with the perturbed c-i cloud. The direct interaction energy is:
\begin{equation}
\delta{\cal F}_{int}=2\tau\log{{\kappa(\lambda+h)\over 2}}
\label{fint}
\end{equation}
The solution of the perturbed PB equation
Eq.(\ref{PPB}) leads to the ``self-energy'':
$\delta{\cal F}_{self}=1/2 l_B \tau^2
{\cal I}_p$, with
\begin{eqnarray}
{\cal I}_p=\int_0^{p_M} dp {\left(1+p(1+\bar h)\right)^2\over
p^3(1+\bar h)^2}\times\cr
\left(e^{-2p\bar h}{1-p+p^2\over 1+p+p^2}-
{1-p(1+\bar h)\over 1+p(1+\bar h)}\right)
\label{phih}
\end{eqnarray}
with $p\equiv q\lambda$ and $\bar h\equiv h/\lambda$. The
integral cutoff is $p_M\equiv2\pi\lambda/a$ where $a$ is a
``microscopic'' size of order the rod radius ($a=20A$ for
DNA\cite{alberts}). This complicated expression can be approximated
in the two important limits: for $h\gg\lambda$
\begin{equation}
\delta{\cal F}_{self}\simeq {1\over 2} l_B\tau^2\left({2\over
3}+\log\left({4\pi\over 3}{h\over a}\right)\right)
\label{selfa}
\end{equation}
and for $h\ll\lambda$.
\begin{eqnarray}
\delta{\cal F}_{self}\simeq {1\over 2} l_B \tau^2\left(-\Gamma+{2\pi\over
3\sqrt{3}}+\log{\pi \lambda^2\over a h}\right)
\label{selfb}
\end{eqnarray}
The self-energy behaves very differently for large and small distances from the wall. 
It shows an attraction between the rod and the plane at large distances (the energy 
increases with the distance), which superimposes to the bare attraction between the 
two oppositely charged macroions. At short distances, there is a strong (logarithmically 
divergent) repulsion between the rod and the plane (Eq.(\ref{selfb})). 

The self-energy shows a deep minimum for a position of the rod which only
depends upon the G-C length: $h_{min}\simeq
0.8\lambda$. This minimum results from a balance between the repulsive
image charge effect and the attraction due to the screening of the
c-i in the Gouy-Chapman layer. The self-energy of a cylinder in a bath of mobile charges is
of order $l_B\tau^2\log{L_\kappa/a}$, where $L_\kappa$ is the characteristic
(``screening'') length of the bath. This energy can be interpreted as
the interaction free energy of the charges on the cylinder (the interaction
having a range $L_\kappa$). Note that
although the free ions are mostly of the same sign as the rod, one
can speak of a
``screening'' effect, as the interaction between the rod and the
unperturbed ion cloud
is taken into account in $\delta{\cal F}_{int}$ (Eq.(\ref{fint})). In addition to the 
self-interaction, one should consider the interaction with the image charge, which in 
the case $\epsilon_{z<0}\ll\epsilon_{z>0}$, is a virtual rod of same charge
located at $z=-h$\cite{jackson}. This gives an extra contribution $\sim
l_B\tau^2\log{L_\kappa/h}$ per unit length. The large and short distance behaviors 
Eq.(\ref{selfa},\ref{selfb}) can now be explained from the counterion profile Eq.(\ref{GC}),
 since $L_\kappa\sim h$ for $h>\lambda$ and $L_\kappa\sim \lambda$ for for $h<\lambda$. 

The adsorption of a polyelectrolyte onto a biomembrane or other fluid membranes, which are
 generally a mixture of charged and neutral lipid molecules, involves the movement of 
 surface charges as a response 
to the field created by the rod. We address this case, disregarding 
the fact that the lipid bilayer is a flexible
object which would, to a certain extent, wrap around the PE\cite{dan} - this
interesting phenomenon will be studied in future works. We also
assume that the charge reorganization at the interface is not limited
by the availability of moving charges on the plane. The charge on the plane 
follows a Boltzmann law: $\sigma=\sigma_0e^{-\delta\phi}$. As a result, the boundary condition
for the perturbed field on the plane (corresponding
to Eq.(\ref{PPB})) has to be modified: $\partial_z\delta\tilde\phi|_{z=0}=4\pi
l_B\sigma_0\delta\tilde\phi|_{z=0}$. This affects the self-energy term only, which can
be expressed similarly to Eq.(\ref{phih}): $\delta{\cal F}_{self}=1/2
l_B \tau{}^2 {\cal I}^{bis}_p$ with
\begin{eqnarray}
{\cal I}^{bis}_p=\int_0^{p_M} dp
{\left(1+p(1+\bar h)\right)^2\over p^3(1+\bar
h)^2}\times\cr
\left(e^{-2p\bar h}{3-3p+p^2\over 3+3p+p^2}-
{1-p(1+\bar h)\over 1+p(1+\bar h)}\right)
\label{phih2}
\end{eqnarray}

For $\bar h>1$ this self-energy is equivalent to the constant surface
charge case (Eq.(\ref{selfa})), while for $\bar h<1$, the
approximation gives:
\begin{equation}
\delta{\cal F}_{self}\simeq {1\over 2} l_B \tau^2\left(-\Gamma-{\pi\over
3\sqrt{3}}+\log{\pi \lambda^2\over 6a h}\right)
\label{self2b}
\end{equation}

The mobile (positive) charges of the surface are attracted toward the
rod ($x=0$); this effectively decreases the ``Gouy-Chapman'' screening
length around the
PE and reduces the image charge effect. The case of an annealed surface charge is
qualitatively similar to (see scaling in Eq.(\ref{self2b})), but
quantitatively different from the case of a quenched surface charge.
Since the repulsion of the wall is weakened, the minimum of the 
self-energy is much deeper, and closer to the wall in the case of
moving surface charges.

We now discuss the variation of the line charge of an annealed, or weak, 
polyelectrolyte near a charged wall. The charges on the chain result from a partial
ionization of specific
chemical groups. The ionization occurs at chemical equilibrium with
the free ions in solution, and is governed by quantities such as the
pH of the solution\cite{barrat}. Formally, the PE charge density can
be determined by equating the chemical potential of the charges on
the chain, to a (given) chemical potential $\mu_0$ for the free
charges. For an infinitely long rigid PE in a salt solution, the free
energy (per unit length) of the charges on the chain is the sum of the translational 
entropy of the charges along the rod, the electrostatic energy, and the chemical 
potential: ${\cal F}=\tau
\log{\tau a/e}+l_B\tau^2\log{\kappa a}-\mu_0\tau$. The equilibrium
charge density of the rod for a given chemical potential is obtained
by differentiation of the free energy: $\mu_0=\log{\tau
a}+2l_B\tau\log{\kappa a}$.
The equivalent expression for a charged rod near an oppositely
charged plane can be computed from the electrostatic free
energy Eq.(\ref{fint},\ref{phih}):
$\mu_0=\log{\tau a}+2\pi l_B\tau\ {\cal
I}_p+2\log{\kappa(\lambda+h)/2}$. Because of the minimum in the self-energy, the 
equilibrium charge density $\tau$ is maximum for a finite height of order the G-C 
length, and decreases sharply near the wall. This non-trivial behavior of the charge 
density of an annealed PE near a wall is mostly due to the importance of the image 
charge effect in the vicinity of the wall. It is extended further below in the case 
of highly charge PE with Manning condensation. Note that the optimum charge of the 
weak PE can reach higher values if the charges on the surfaces are mobile, but still
 decreases at shorter distances. 

The perturbative treatment is expected to fail in the important case
of the release of the counterions condensed onto a
highly charged rod, as the rod approaches an oppositely charged
plane. However, the qualitative argument which translates the
concentration of free charges into a local screening length should
still hold in this case.

A charged cylinder surrounded by its
counterions
undergoes the so-called Manning condensation\cite{manning}.
Solutions
of the poisson-Boltzmann equation in this geometry\cite{fuoss}
predict that
if the rod is highly charged (namely $l_B\tau>1$), a finite fraction
$1-\beta$
of the counterions are confined in the close vicinity of the rod. The
electrostatic properties at large distances from the rod are the same
as those of a rod with an effective charge  $l_B\tau^*\simeq 1$. The
(over)simplified Oosawa
picture of  counterion condensation\cite{oosawa1} gives a
qualitative account of
this phenomenon\cite{holm}; it is based on a chemical equilibrium between two
types of counterions:
condensed c-i with a reduced entropy and subjected to a large
electrostatic attraction from the rod on the one hand, and free c-i in solution far
from the rod on the other hand. This picture can be adapted to the case of a cylinder
of charge $\tau$ in a salt solution of density $n_0$ (or screening
length $\kappa^{-1}$). The effective charge of the
cylinder, $\tau^*=\beta\tau$, is obtained by balancing the chemical
potentials of the condensed c-i $\mu_{cond}=\log\left[{(1-\beta)\tau
v\over\pi a^2}\right]+2 l_B\tau\beta\log\left[\kappa a\right]$ and of
the c-i dispersed among the salt molecules
$\mu_{free}=\log\left[n_0v\right]$ ($v$ is the volume of a c-i
molecule). The resulting fraction of free c-i $\beta$ is: $\log\left[8 l_B\tau(1-\beta)\right]=\left(1-l_B\tau\beta\right)
\log\left[(\kappa a)^2\right]$. This result is very similar to the Oosawa 
relationship for a rod which would occupy a (small) volume fraction 
$\Phi=(\kappa a)^2$ in a salt-free
solution\cite{oosawa}, namely $\beta\simeq1$ for $l_B\tau<1$ and
$\beta l_B\tau\simeq 1$ for $l_B\tau>1$.

The counterion condensation on a rod near a charged plane can be
derived in the same way. The chemical potential of the condensed
counterions depends upon the electrostatic potential on the rod,
which can be determined (at the level of the scaling laws) using the
calculation of the previous sections. Equating the chemical
potentials of the free and condensed c-i, we obtain
the fraction of free c-i $\beta$:
\begin{eqnarray}
\log\left[8l_B\tau(1-\beta)\right]=\log\left[(\kappa
a)^2\right]-\cr2l_B\tau\beta\log\left[{a h\over
L_\kappa(h)^2}\right]+2\log\left[{\kappa(h+\lambda)\over 2}\right]
\label{taufin}
\end{eqnarray}
where the local screening length follows the asymptotic behaviors:
$L_\kappa=\lambda$ for $0<h<\lambda$ and
$L_\kappa=h$
for $\lambda<h<\kappa^{-1}$. Examples of the counterion release as a function of 
the distance to
the wall is shown on Fig.1  (for which proper screening due to the salt has been 
taken into account for $\kappa h>1$). The Manning parameter $l_B\tau=4$ and the 
radius
$a=20 A$ are of order those for DNA, three surface charge densities ranging from 
$\lambda=200 A$ to $50A$ as been considered, corresponding to one charge every 
$(100 A)^2$ to $(50A)^2$. The salt concentration $\kappa^{-1}=1000 A$ corresponds
 to a
concentration of $10^{-5} mol/l$. While the c-i
are released as the rod penetrates the Gouy-Chapmann layer, full
release ($\beta=1$) is only reached for $\lambda=50A$ - the maximum being $60\%$ 
for $\lambda=200A$ (it is of the order of $30\%$ for the free rod).
Furthermore, the free charges recondense at short distance if $\lambda$ is large 
enough; the effective charge reaches
$50\%$ of the bare charge for $\lambda=200A$. It can be shown that for large salt
concentration or weak surface charge of the wall: $\kappa^2\lambda^3<2 a$, the 
PE in contact with the wall can have a lower effective charge than the free PE.

It should be noted that the short range repulsion
due to the image charge can prevent real adsorption with an equilibrium
contact between the rod and the plane. In many cases, however, a short range
attraction of non electrostatic origin (such as a hydrophobic force)
dominates the adsorption. This interaction must be added to the
electrostatic free energy calculated here in order to study
the equilibrium adsorption.

Different conclusions are reached in the case of moving surface
charges. A highly charge rod has a strong effect on the surface
charge distribution, for the potential it creates is likely to dominate over the 
Gouy-Chapman potential, even near the wall. A quantitative
description of the phenomenon would require the solution of the
full non-linear Poisson-Boltzmann equation, with complex (non-linear as
well) boundary conditions. In the following, we merely try to give a
feeling of the way moving charges can influence the counterion
release of an adsorbed polyelectrolyte. We assume that the surface charge 
distribution obeys Boltzmann
statistics: $\sigma_{x=0}=\sigma_0  e^{-\phi_h}$, where the
potential $\phi_h$ created by the rod at the surface reflects the
screening due to the Gouy-Chapmann layer, the image charge, and the
fraction $(1-\beta)$ of condensed counterions: $\phi_h\sim-
2l_B\tau\beta\log{(L_\kappa/h)^2}$. Since the local screening length
$L_\kappa \simeq(\lambda+h)/\sqrt{2}$ is influenced by the surface
charge $\sigma$, which in turn, depends upon the screening length, we
obtain a self-consistent relationship for the local
Gouy-Chapman length near the rod:
\begin{equation}
\lambda(h)\left({h+\lambda(h)\over \sqrt{2} h}\right)^{4
l_B\tau\beta}=\lambda_0
\label{localGC}
\end{equation}
This expression shows that moving surface charges strongly reduce the
screening length near the rod, which becomes of order $h$ when the
rod is close to the wall (recall that $l_B\tau\beta\simeq1$). As a
consequence, the interaction with the image charge and the self-interaction 
along the rod are both strongly screened. The dominant interaction
between the cylinder and the wall is the attractive part given by
Eq.(\ref{fint}), and we are likely to observe a full release of the
condensed counterions in this case.

To summarize, we have studied the evolution of the effective charge of a
polyelectrolyte near an oppositely charged plane. The
case of a weak PE is studied fairly rigorously, via a perturbative
treatment of the non-linear Poisson-Boltzmann equation. We show that
at large distance, the charge of the PE increases as the distance to
the wall $h$ decreases, as expected. However, the charge decreases
``strongly'' as the PE enters the Gouy-Chapman layer
($h\simeq\lambda$) because of the combined effect of image charge
(for the most common case of a wall with a low dielectric constant)
and self-interaction along the PE. These two effects are very much
influenced by the value of the Gouy-Chapman length $\lambda$, and are
partly suppressed in the case of a fluid interface with moving surface charges 
(for a fluid lipid bilayer for instance), where the
charge of the adsorbed PE can reach higher values.

The most interesting case of a strongly charge PE beyond the Manning
condensation threshold (such as DNA) is discussed qualitatively,
using ``scaling'' arguments inferred from the perturbation theory. We
predict that in the case of fixed surface charges, and in contrary to
a widely spread idea, most of the condensed counterions \underline{are not 
released} if the Gouy-Chapman length is larger than the
radius of the rod: $\lambda\gg a$. In the case of freely moving
surface charges, a full release of the condensed counterions is
expected, as the effective Gouy-Chapman length near the rod is of
order the rod radius.

In all the discussion, we have assumed that the Gouy Chapman length is
larger than a molecular size. In many real cases, the two lengths are
of the same order of magnitude and the finite size of the ions must
be taken into account in order to obtain quantitative results. In this
case our results can at best be considered as qualitative.

We would like to thank A. Johner and R. Netz for many stimulating discussions.

\newpage
\begin{figure}[h]
\vskip2truecm
\centerline{ \epsfxsize=10truecm \epsfbox{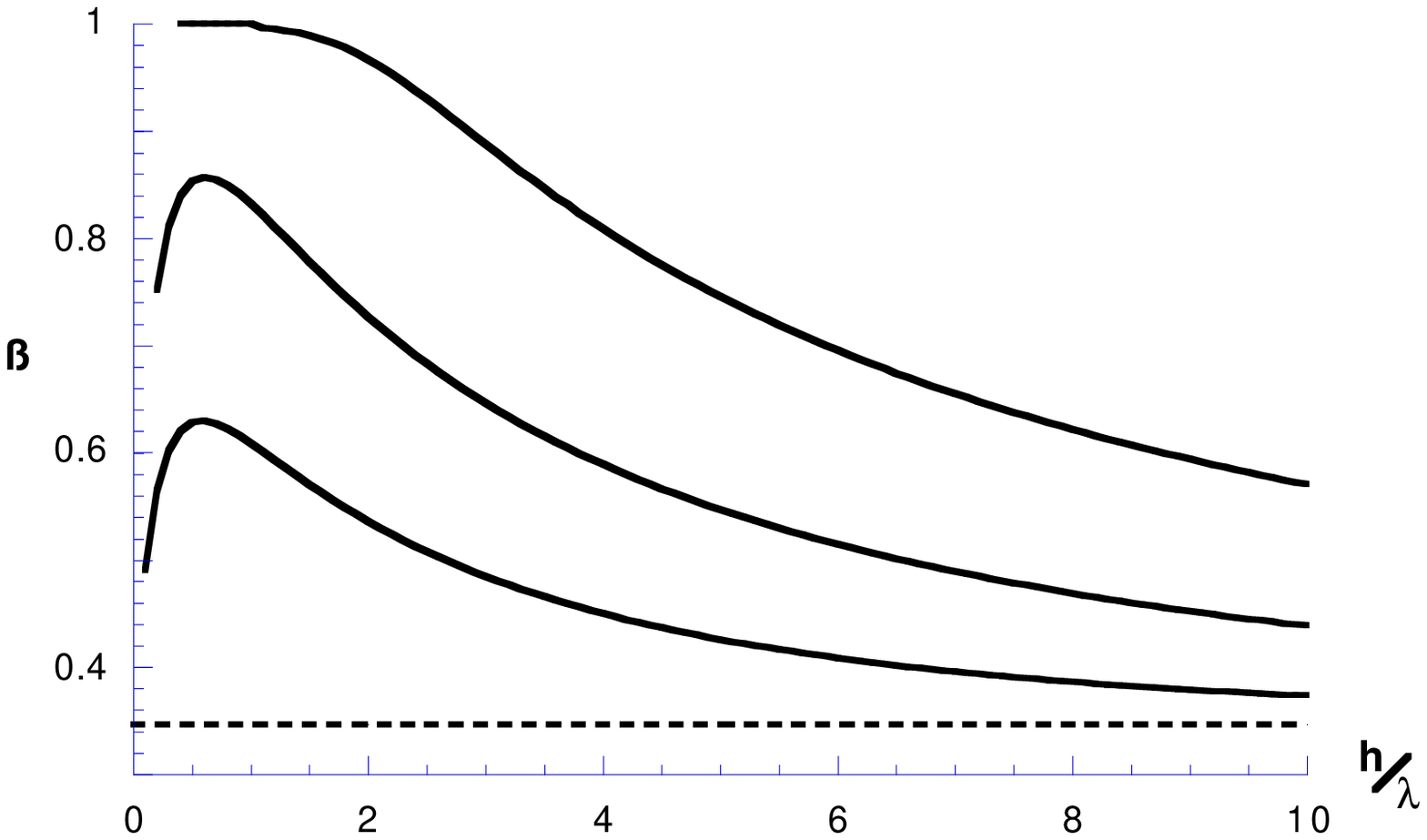} }
\vskip2truecm
\caption{\protect\small Fig.1 Fraction $\beta$ of free counterions as a
function of the distance to the wall in $\lambda$ units (dashed:
$\beta(h\rightarrow\infty))$ for $a=20A$, $\l_B\tau=4$, $\kappa^{-1}=1000A$, for decreasing values of the Gouy-Chapmann length: $\lambda=200A$, $100A$ and $50A$}
\end{figure}

\end{document}